\documentclass[aps,twocolumn,prb,showpacs,superscriptaddress,floatfix]{revtex4}
\usepackage{amsmath} \usepackage{graphicx} \usepackage{amssymb}
\usepackage{bm}

\newcommand{\e}{\mathrm{e}} 
\newcommand{\Tc}{T_\mathrm{c}}

\DeclareMathAlphabet{\mathitbf}{T1}{cmr}{bx}{it} 

\begin{document}
  
\title{The Spin Glass Phase in the Four-State, Three-Dimensional Potts
  Model}

\author{A.~Cruz} \affiliation{Departamento
  de F\'\i{}sica Te\'orica, Universidad
  de Zaragoza, 50009 Zaragoza, Spain.} 
  \affiliation{Instituto de Biocomputaci\'on y
  F\'{\i}sica de Sistemas Complejos (BIFI), Zaragoza, Spain.}

\author{L.A.~Fernandez} \affiliation{Departamento
  de F\'\i{}sica Te\'orica I, Universidad
  Complutense, 28040 Madrid, Spain.} 
  \affiliation{Instituto de Biocomputaci\'on y
  F\'{\i}sica de Sistemas Complejos (BIFI), Zaragoza, Spain.}

\author{A.~Gordillo-Guerrero}\affiliation{Dpto. de Ingenier\'{\i}a
  El\'ectrica, Electr\'onica y Autom\'atica,  Universidad de Extremadura.
  Avda. de la Universidad s/n. 10071. C\'aceres, Spain.}
\affiliation{Instituto de Biocomputaci\'on y F\'{\i}sica de Sistemas
  Complejos (BIFI), Zaragoza, Spain.}

\author{M.~Guidetti} \affiliation{Dipartimento
  di Fisica Universit\`a di Ferrara and INFN - Sezione di Ferrara,
  Ferrara, Italy.} 

\author{A.~Maiorano} \affiliation{Dipartimento
  di Fisica Universit\`a di Ferrara and INFN - Sezione di Ferrara,
  Ferrara, Italy.}  \affiliation{Instituto de Biocomputaci\'on y
  F\'{\i}sica de Sistemas Complejos (BIFI), Zaragoza, Spain.}

\author{F.~Mantovani} \affiliation{Dipartimento
  di Fisica Universit\`a di Ferrara and INFN - Sezione di Ferrara,
  Ferrara, Italy.} 

\author{E.~Marinari} \affiliation{Dipartimento di Fisica, INFN and
  INFN, Universit\`a di Roma ``La Sapienza'', 00185 Roma, Italy.}

\author{V.~Martin-Mayor} \affiliation{Departamento de F\'\i{}sica
  Te\'orica I, Universidad Complutense, 28040 Madrid, Spain.} 
\affiliation{Instituto de Biocomputaci\'on y
  F\'{\i}sica de Sistemas Complejos (BIFI), Zaragoza, Spain.}

\author{A.~Mu\~noz Sudupe} \affiliation{Departamento
  de F\'\i{}sica Te\'orica I, Universidad
  Complutense, 28040 Madrid, Spain.} 

\author{D.~Navarro} \affiliation{Departamento  de Ingenier\'{\i}a,
  Electr\'onica y Comunicaciones and Instituto de Investigaci\'on en\\
  Ingenier\'{\i}a de Arag\'on (I3A), Universidad de Zaragoza, 50018 Zaragoza, Spain.}

\author{G.~Parisi} \affiliation{Dipartimento di Fisica, INFN and
  INFN, Universit\`a di Roma ``La Sapienza'', 00185 Roma, Italy.}

\author{S.~Perez-Gaviro} \affiliation{Instituto de Biocomputaci\'on y
  F\'{\i}sica de Sistemas Complejos (BIFI), Zaragoza, Spain.}

\author{J.~J.~Ruiz-Lorenzo} \affiliation{Departamento de
  F\'{\i}sica, Universidad de Extremadura, 06071 Badajoz, Spain.}
\affiliation{Instituto de Biocomputaci\'on y
  F\'{\i}sica de Sistemas Complejos (BIFI), Zaragoza, Spain.} 

\author{S.F.~Schifano} \affiliation{Dipartimento
  di Fisica Universit\`a di Ferrara and INFN - Sezione di Ferrara,
  Ferrara, Italy.} 

\author{D.~Sciretti} \affiliation{Departamento
  de F\'\i{}sica Te\'orica, Universidad
  de Zaragoza, 50009 Zaragoza, Spain.}
  \affiliation{Instituto de Biocomputaci\'on y
  F\'{\i}sica de Sistemas Complejos (BIFI), Zaragoza, Spain.}

\author{A.~Tarancon} \affiliation{Departamento
  de F\'\i{}sica Te\'orica, Universidad
  de Zaragoza, 50009 Zaragoza, Spain.} 
  \affiliation{Instituto de Biocomputaci\'on y
  F\'{\i}sica de Sistemas Complejos (BIFI), Zaragoza, Spain.}

\author{R.~Tripiccione} \affiliation{Dipartimento
  di Fisica Universit\`a di Ferrara and INFN - Sezione di Ferrara,
  Ferrara, Italy.} 
 
\author{J.L.~Velasco} \affiliation{Instituto de Biocomputaci\'on y
  F\'{\i}sica de Sistemas Complejos (BIFI), Zaragoza, Spain.}

\author{D.~Yllanes}  \affiliation{Departamento de F\'\i{}sica
  Te\'orica I, Universidad Complutense, 28040 Madrid, Spain.}
  \affiliation{Instituto de Biocomputaci\'on y
  F\'{\i}sica de Sistemas Complejos (BIFI), Zaragoza, Spain.}

\author{A.P.~Young} \affiliation{Department of Physics,  
  University of California, Santa Cruz, CA 95064, USA }

\date{\today}

\begin{abstract}
  We perform numerical simulations, including parallel tempering, on the Potts
  glass model with binary random quenched couplings using the JANUS
  application-oriented computer.  We find and characterize a glassy
  transition, estimating the location of the transition and the value of the
  critical exponents. We show that there is no ferromagnetic transition in a
  large temperature range around the glassy critical temperature. We also
  compare our results with those obtained recently on the ``random
  permutation'' Potts glass.
\end{abstract}


\maketitle

%%%%%%%%%%%%%%%%%%%%%%%%%%%%%%%%%%%%%%%%%%%%%%%%%%%%%%%%%%%%%%%%%%
\section{Introduction}

Potts models are among the building blocks of statistical mechanics,
and their disordered versions (Potts glasses)
are commonly used to
describe a large class of anisotropic orientational
glasses.\cite{BINDER}  For example, if a crystal of molecular nitrogen
is disordered by including some percentage of argon, the resulting
compound,
Ar$_{1-x}$(N$_2$)$_x$, is a disordered
quadrupolar glass.\cite{EXP} The Potts glass is one of the models of
choice to describe materials of this type.

The four state ($p=4$) pure Potts model in two dimensions ($D=2$) describes 
the adsorption of N$_2$ on Kr in graphite
layers.\cite{ABSOR_DOMANY} In $D=3$ it describes the behavior of FCC
antiferromagnetic materials (NdSb, NdAs and CeAs for example)
where the magnetic field points in the $<$1,1,1$>$ direction\cite{FCC_DOMANY}.
The site diluted version of the Potts
model can describe, for example, the adsorption of hydrogen on the
$(1,1,1)$ plane of nickel which has been previously disordered by
inserting oxygen atoms.\cite{POTTS2D_EXP}  In this paper we will study
the four state glassy Potts model, in which quenched random disorder
induces frustration. This model presents at least three interesting
theoretical problems that are still unsolved.

The \textit{first} of these
is the nature of the spin glass
phase transition; one needs to reliably compute the critical
temperature and the critical exponents in order to characterize the
\textit{universality class} of the model.

The \textit{second} issue is how the qualitative features of phase diagram,
including
spin glass and ferromagnetic phases,
evolve when going from the mean field models to realistic, finite
dimensional models.
For example,
previous work \cite{ELDER1,GROSS,DeSANTIS} has shown
that, at low temperatures, the standard Potts glass (in which the Potts
coupling between two spins can have positive or negative sign) develops
spontaneous ferromagnetic ordering, which can affect, or even prevent,
a spin glass
phase transition. Furthermore, in mean field theory the value of this
ferromagnetic transition temperature, $T_\mathrm{FM}$, varies with the number
$p$ of Potts states as $T_{\mathrm{FM}}=(p-2)/2$, which gives $T_{\mathrm
{FM}}=1$ for the four state model. Mean field analysis also shows that for
$p\le 4$ the spin glass transition temperature (where replica symmetry gets
broken) is $T_\mathrm{RSB}=1$: if $p=4$ the two transition temperatures
coincide.  Hence an interesting open problem is to check whether or not this
result, valid for $D=\infty$, also holds when the dimensionality is finite.
Also relevant here is that,
in the mean field picture, the $p=4$ glassy model is
``marginal'', since for $p\le 4$ the transition is continuous whereas for $p>
4$ the order parameter $q(x)$ is discontinuous (even if, as usual in spin
glasses, there is no latent heat).\cite{GROSS} In a series of interesting
papers Binder, Brangian and Kob\cite{BBK} (see also the recent work by
Katzgraber, Lee and Young\cite{KLY}) study the ten state glassy Potts model,
and find that for such high number of states the mean field and the finite
dimensional cases \textit{are very different}.
Here we investigate whether the same is
true for $p=4$.

The \textit{third} relevant issue is again related to
universality. In order to avoid a possible contamination of the spin
glass transition point by the effects of the ferromagnetic phase,
Marinari, Mossa and Parisi\cite{GLASSY-POTTS} have introduced a new
class of
glassy Potts models, the ``random permutation'' Potts glass,
where a gauge symmetry protects the model
against a ferromagnetic transition.
This approach has the advantage of being 
closer to reality since, in real quadrupolar
glasses, ferromagnetism plays no role.
One of these models has been thoroughly studied recently by some of
the authors of the present work,\cite{POTTS-PERM}
and its behavior found
to be consistent with a
Kosterlitz-Thouless phase transition.
One of its signatures is that, below the critical point,
data for
the correlation length divided by lattice size,
$\xi/L$,
for different sizes 
merge into a single curve.
However, given the precision of the numerical data and allowing for
corrections to finite-size scaling, one cannot exclude a
value of the lower critical behavior near
and slightly below $D=3$.

A further motivation for this study is to investigate how the behavior of the
Potts glass changes with $p$.
For $p=3$, a
Potts glass transition occurs \cite{KLY} with
critical exponents $\nu \simeq 1.2$ and $\eta \simeq 0.02$, while,
for $p=10$, Ref.~\onlinecite{KLY} finds no phase transition in agreement
with Ref.~[\onlinecite{BBK}].  In addition, we note
the Ising spin glass model, which corresponds to $p=2$,
has $\nu \simeq 2.5$ and $\eta \simeq -0.4$ (see
Ref.~(\onlinecite{PELI,KKY})).
It would therefore be very interesting to get a
consistent picture of how
the nature of the Potts glass transition evolves with the
number of states $p$.

In an attempt to give reliable answers to these questions, we
have performed extensive numerical simulations using one unit
of the JANUS dedicated computer (which has a total of $16$
units).\cite{IANUS2} We have been able to thermalize
the $p=4$ Potts glass model on an
$L=16$ cubic lattice down to the low temperature phase: 
this gives us information on far larger
lattice sizes
than had been possible before.

The outline of the paper is the following. In section $2$ we
introduce our model and physical observables. In section $3$ we
describe the numerical methods that we have used in the simulations.
In section $4$ we describe our tests of thermalization and our
approach to data analysis, and we analyze our findings, both for the
overlap and for the magnetization.  The main results are that we have
been able to characterize the spin glass transition and that no onset of
ferromagnetic order has been found at and below the spin glass
transition point. We discuss these findings in section $5$.

%%%%%%%%%%%%%%%%%%%%%%%%%%%%%%%%%%%%%%%%%%%%%%%%%%%%%%%%%%%%%%%%%%
\section{Model and observables}\label{sec:MODEL}

In the $p$-state Potts model, each site $i$ of a three dimensional
cubic lattice of linear size $L$ with periodic boundary conditions
has a 
scalar spin ${s_i}$ which takes one of the values
$1,2, \dots , p$.
The Hamiltonian of the standard Potts glass model is
\begin{equation}
  \label{eq:ham}
  H = - \sum_{\langle i,j \rangle}J_{ij}\, \delta_{s_i,s_j}\,,
\end{equation}
where the sum runs over all nearest neighboring sites.  Two
neighboring sites $i$ and $j$ interact with energy $-J_{ij}$ when their
spin states $s_i$ and $s_j$ have the same value, and 
otherwise their energy is
zero.  The couplings
$J_{ij}$ are independent quenched random variables taken from a
bimodal distribution ($J_{ij}=\pm 1$) with zero average.  From now on
we will focus on the four-state ($p=4$) case.

It is possible to 
represent the state of site $i$ by a $(p-1)$-dimensional vector,
$\mathitbf{S}_i$, equal to one of the $p$
unit vectors $\mathitbf{S}_a$
pointing to the corners of a hyper-tetrahedron in $(p-1)$-dimensional
space.
These vectors satisfy the
relations
\begin{equation}
  \label{eq:simplex}
  \mathitbf{S}_a \cdot \mathitbf{S}_b = \frac{p \, \delta_{a b}-1}{p-1}\,.
\end{equation}
Equation
\eqref{eq:simplex} defines the \textit{simplex} representation, that
we will use to describe the observables measured in the simulations.

In order to investigate the possible presence of (spurious)
ferromagnetic effects we have carefully checked the value of the
magnetization, looking for possible signs of spontaneous ferromagnetic
ordering.
In the simplex representation we define the
vector magnetization as
\begin{equation}
  \label{eq:magnet}
  \mathitbf{m} = \frac{1}{N} \sum_{i=1}^N \mathitbf{S}_i \,,
\end{equation}
where $N\equiv L^3$ is the number of spins.

The existence of a possible transition to a ferromagnetic phase has
been also analyzed by studying the magnetic susceptibility
\begin{equation}
  \label{eq:chifm}
  \chi_{\mathrm{M}} = N\, \overline{\langle | \mathitbf{m}|^2 \rangle}\,,
\end{equation}
where $\langle (\cdot\cdot \cdot) \rangle$ stands for the thermal average and
$\overline{(\cdot\cdot\cdot)}$ denotes the disorder average.

To study the glass transition we define the spin-glass order parameter
as a tensorial overlap between two replicas (i.e. two copies of the
system defined with 
identical couplings whose spin values evolve independently).  The
standard definition of its Fourier transform with wave vector
$\mathitbf{k}$ is \cite{KLY}
\begin{equation}
  \label{eq:overlap}
  q^{\mu \nu}(\mathitbf{k}) = \frac{1}{N} \sum_i 
  S_{i}^{(1)\mu} S_{i}^{(2)\nu} \e^{i \mathitbf{k} \cdot \mathitbf{R}_i}\,,
\end{equation}
where $S_{i}^{(1)\mu}$ is the $\mu$-th component of the spin
of the first replica in the simplex representation, and
$S_{i}^{(2)\nu}$ is the $\nu$-th component of the spin in the
second replica.

The momentum-space, spin-glass susceptibility is defined by
\begin{equation}
  \label{eq:chisg}
  \chi_{q}(\mathitbf{k})= N \sum_{\mu,\nu}
  \overline{\langle |q^{\mu \nu} (\mathitbf{k})|^2 \rangle}\,.
\end{equation}
We also define the correlation length $\xi$ in terms of the Fourier
transform\cite{VICTORAMIT} in Eq.~(\ref{eq:chisg}) as
\begin{equation}
  \label{eq:xi}
  \xi = \frac{1}{2\sin{(\mathitbf{k}_{\mathrm{m}}/2)}} 
  \bigg( \frac{\chi_q(0)}{\chi_q(\mathitbf{k}_{\mathrm{m}})} 
  - 1 \bigg)^{1/2}\,,
\end{equation}
where $\mathitbf{k_{\mathrm{m}}}$ is the minimum wave vector allowed
within the lattice. Periodic boundary conditions imply that this
vector is $\mathitbf{k_{\mathrm{m}}}=(2 \pi /L,0,0)$ or one of the
two other related vectors in which the components are permuted.
The definition in Eq.~(\ref{eq:xi}) arises naturally 
on a finite lattice.

We will base a large part of our analysis on the dimensionless
correlation length $\xi/L$, i.e. on the correlation length 
divided by the lattice size.  This quantity does
not depend on $L$ (asymptotically for large $L$) at the transition
temperature, which allows us to obtain a precise estimate of $\Tc$ 
from the value of $T$ at which data for 
different lattice sizes cross.\cite{VICTORAMIT}

%%%%%%%%%%%%%%%%%%%%%%%%%%%%%%%%%%%%%%%%%%%%%%%%%%%%%%%%%%%%%%%
\section{Numerical methods}\label{sec:MONTECARLO}

We have simulated three dimensional cubic lattices with linear sizes
$L=4$, $6$, $8$ and $16$. Because spin-glass simulations  have
very long relaxation times, we used the parallel
tempering (PT) algorithm \cite{PT} to speed up the dynamical
process that brings the system to thermal equilibrium and
eventually explores it.
Physical quantities are only measured after the system has been
brought to equilibrium.

The dynamics is comprised of single-spin updates and temperature
swaps.
The single-spin updates are carried out with a
sequential heat bath (HB) algorithm.
We define a \textit{Monte Carlo
sweep} (MCS) as $N$ sequential trial updates of the HB algorithm
(i.e. every spin undergoes a trial update once).

The PT algorithm (applied to a given realization of the quenched
disorder, that we will call a sample) is based on simulating a number
of copies of the system with different values of the temperature
but the same interactions.
Exchanging the temperature of two copies with adjacent temperatures with a
probability that respects the
detailed balance condition
is the crucial mechanism of PT.
The result is that each copy of the system drifts
in the whole allowed temperature range (that has been decided a
priori). 
When a copy is at a high temperature it equilibrates fast and so each time it
descends to low temperature it is likely to be in a different valley in the
energy landscape.

The HB and the PT algorithms require high quality
random numbers; we generate them with a $32$-bit Parisi-Rapuano shift
register \cite{PARISI-RAPUANO} pseudo-random number generator.

Details about our numerical simulations are summarized in Table
\ref{tab:SIM_DETAILS}. The simulation of the smaller lattices, with
$L=4$ and $6$, was performed on standard computers. More powerful
computational resources are needed to deal with the $L=8$ and $16$
systems, so we have studied them on a prototype board of the Janus
\cite{IANUS2} computer, an FPGA based computer optimized for a
relatively small set of hard computational problems (among them, spin
glass simulations). A performance comparison between an Intel(R)
Core2Duo(TM) processor and one Janus processor (one FPGA) shows that the
latter is about one thousand times faster\cite{JANUS08} when simulating
Potts models.  JANUS has allowed us to thermalize a large number of
samples for bigger sizes than would have been feasible on a standard
computer.  The computational effort behind our analysis amounts to
approximately 6 years CPU time on a 2.4 GHz Intel(R) Core2Duo(TM)
processors for $L=8$ and thousands of CPU-years for $L=16$.

Data input and output is a critical issue for JANUS
performance, so we had to carefully choose how often to read
configuration data; in general, we end up taking fewer
measurements than in simulations on a traditional PC.  Having
fewer (but less correlated) measurements does not affect the quality of
our results. We read and analyze values of physical observables
every $2 \times 10^5$ MCS for both $L=8$ and $16$ (see
Table~\ref{tab:SIM_DETAILS} for details).

\begin{table}[h]
  \begin{center}
    % \begin{ruledtabular}
    \begin{tabular}{|c|c|c|c|c|c|c|}\hline
      $L$ & $N_\mathrm{samples}$ & MCS & $[\beta_\mathrm{min},\beta_\mathrm{max}]$ & $~N_{\beta}~$ & $N_\mathrm{HB}$  & $N_m$  \\
      \hline
      \hline
      4  & 1000 & $3.2\times 10^5$      & [2.0,6.0] & 9  & 5  & $10^3$\\
      \hline                                                  
      6  & 1000 & $8\times 10^5$        & [2.5.5.0] & 7  & 5  & $10^3$\\
      \hline
      8  & 1000 & $2\times 10^8$        & [2.7,4.2] & 16 & 10 & $2\times 10^5$\\
      \hline                                                  
      16 & 1000 & $8\times 10^9$   & [1.7,4.1] & 32 & 10 & $2\times 10^5$\\
      \hline
    \end{tabular}
  \end{center}
  % \end{ruledtabular}
  \caption{For each lattice size we show the number of disorder
    samples that we have analyzed, the
    number of MCS per sample, the range of simulated inverse temperatures
    $\beta=1/T$,
    the number of (uniformly distributed) $\beta$ values
    used for PT, the number of MCS performed between two
    PT steps ($N_\mathrm{HB}$), and the number of MCS between
    measurements ($N_m$).}\label{tab:SIM_DETAILS}
\end{table}

On the larger lattices, we perform a PT step every 10 MCS  while on the
smaller lattices this value is 5. In a standard computer the PT algorithm
takes a negligible amount of time, compared to a whole MCS.  However, in JANUS
the clock cycles needed by one PT step are more than those needed for a MCS.
For this reason we chose to increase the  number of MCS between two PT
steps.  However, this number should not be \textit{too} large, as we
do not want to negatively affect the PT efficiency.  A preliminary analysis
has been carried over to test how the PT parameter would affect the simulation
results, and we have selected a value that seems to be well
optimized (see Table~\ref{tab:SIM_DETAILS}).

%%%%%%%%%%%%%%%%%%%%%%%%%%%%%%%%%%%%%%%%%%%%%%%%%%%%%%%%%%%%%%%%%%%%
\section{Results}\label{sec:RESULTS}

%%%%%%%%%%%%%%%%%%%%%%%%%%%%%%%%%%%%%%%%%%%%%%%%%%%%%%%%%%%%%%%%%%
\subsection{Thermalization Tests}\label{sub:THERMALIZATION}

We start by briefly discussing the tests that we 
performed to check if our numerical data are really well thermalized.
We use a standard test in which 
a given physical quantity
is averaged (first over the thermal noise and then over the quenched
disorder) over logarithmically increasing time windows.  
Equilibrium is reached when successive values converge. 
We emphasize that it
is crucial for time to be plotted on logarithmic scale.

\begin{figure}[h]
  \includegraphics[width=0.7\columnwidth,angle=270]{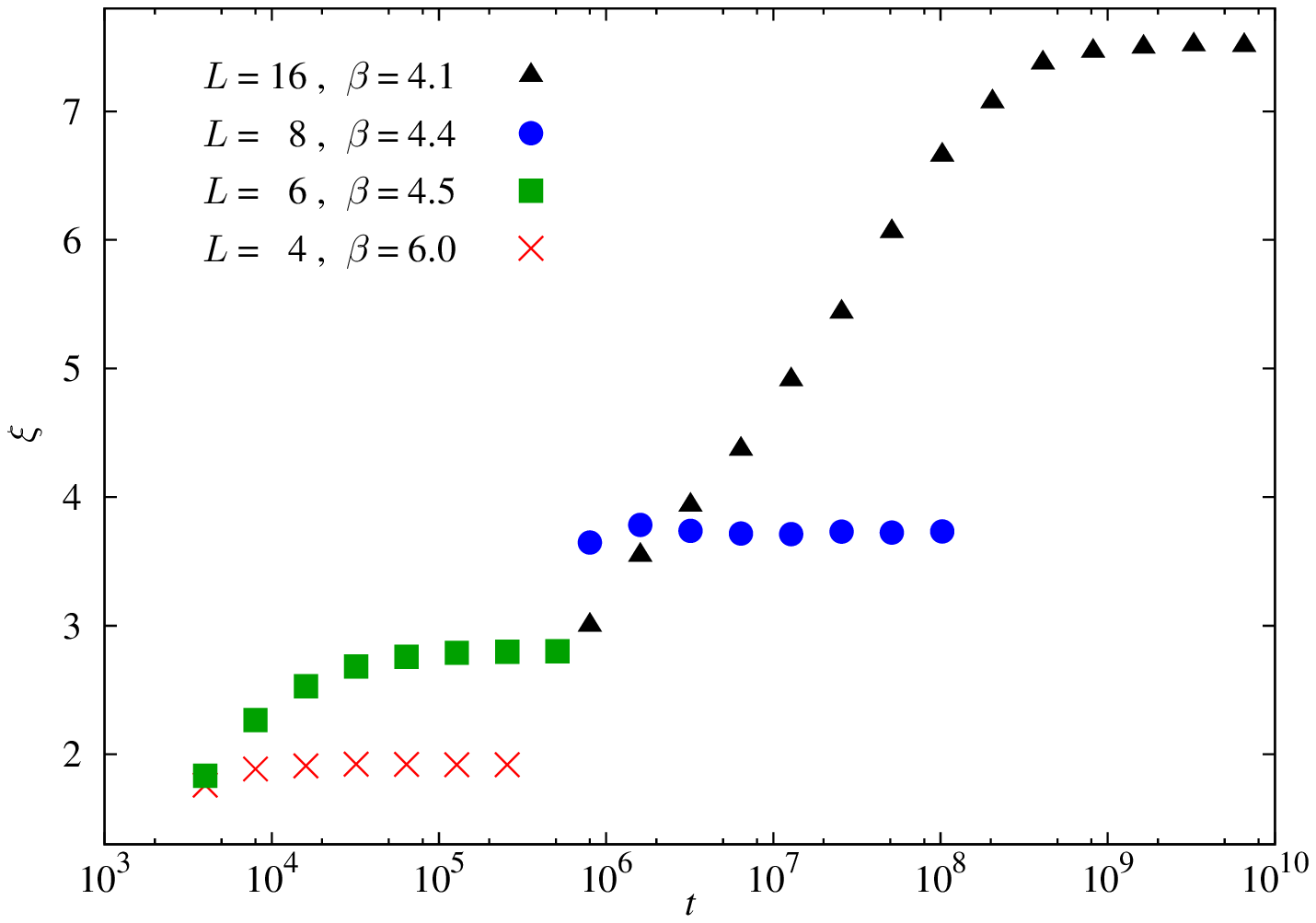}
  \caption{
  A thermalization test. We show the behavior of the
    time dependent spin glass correlation length as a function
    of Monte Carlo time. We have averaged the correlation length
    using a logarithmic binning procedure. We show data for the
    lowest temperature simulated for each size.}
  \label{fig:term}
\end{figure}

We are interested in the correlation length, defined in
Eq.~(\ref{eq:xi}), which is plotted in
Fig. \ref{fig:term} at the lowest simulated temperature (the hardest
case for thermalization). We see that the values of
the correlation length reach a clear plateau for all sizes, strongly
suggesting that our samples have reached thermal equilibrium.  This
analysis also provides useful information about the number of sweeps
that have to be discarded at the beginning of the Monte-Carlo history in
order to use only equilibrated configurations.

%%%%%%%%%%%%%%%%%FIN QUI %%%%%%%%%%%%%%%%%%%%%%%%%%%%%%%%%%%%%%%%%%%%%%%%%%%%%%

\subsection{Finite Size Scaling Analysis; The Quotient Method}\label{sub:QUOTIENT}

To measure the critical exponents we used the quotient
method.\cite{QUOTIENT,VICTORAMIT} 
In this approach one compares results for lattice sizes $L$ and $s L$
for integer $s$ which here we take to be 2.  Firstly, for
a pair of lattice sizes $L$ and $s L$, we find the point, $\beta =
\beta_{\mathrm{cross}}$, where the
correlation length divided by system size is equal for the two sizes,
i.e.
\begin{equation}
\frac{\xi(s L,\beta_{\mathrm{cross}})}{s\, L}=
\frac{\xi(L,\beta_{\mathrm{cross}})}{L}\, ,
  \label{BETANAIVE}
\end{equation}
or equivalently
\begin{equation}
Q_\xi(L, s L) \equiv \frac{\xi(s L,\beta_{\mathrm{cross}})}
{\xi(L,\beta_{\mathrm{cross}})} = s \, .
\end{equation}

We then determine similar ratios for other observables. If an observable
$O$ diverges near the critical
temperature as $t^{-x_O}$, where $t$ is the reduced critical
temperature, then we expect
\begin{equation}
Q_O(L, s L) \equiv \frac{O(s L,\beta_{\mathrm{cross}})}
{O(L,\beta_{\mathrm{cross}})} = s^{x_O/\nu}+O(L^{-\omega})\, ,          
\label{QUO}
\end{equation}
where $\omega$ is the exponent describing the leading corrections to
scaling.

Applying Eq.~(\ref{QUO}) to the operators $\partial_\beta \xi$, and
$\chi_q$ yields respectively the critical exponents $1+1/\nu$ and
$2-\eta_q$. Similarly, if we apply Eq.~(\ref{QUO}) to the magnetic
susceptibility we obtain the exponent $2-\eta_m$.

%%%%%%%%%%%%%%%%%%%%%%%%%%%%%%%%%%%%%%%%%%%%%%%%%%%%%%%%%%%%%%%%%%
\subsection{Overlap Critical Exponents}\label{sub:EXPONENTS}

\begin{figure}[h]
  \includegraphics[width=0.7\columnwidth,angle=270]{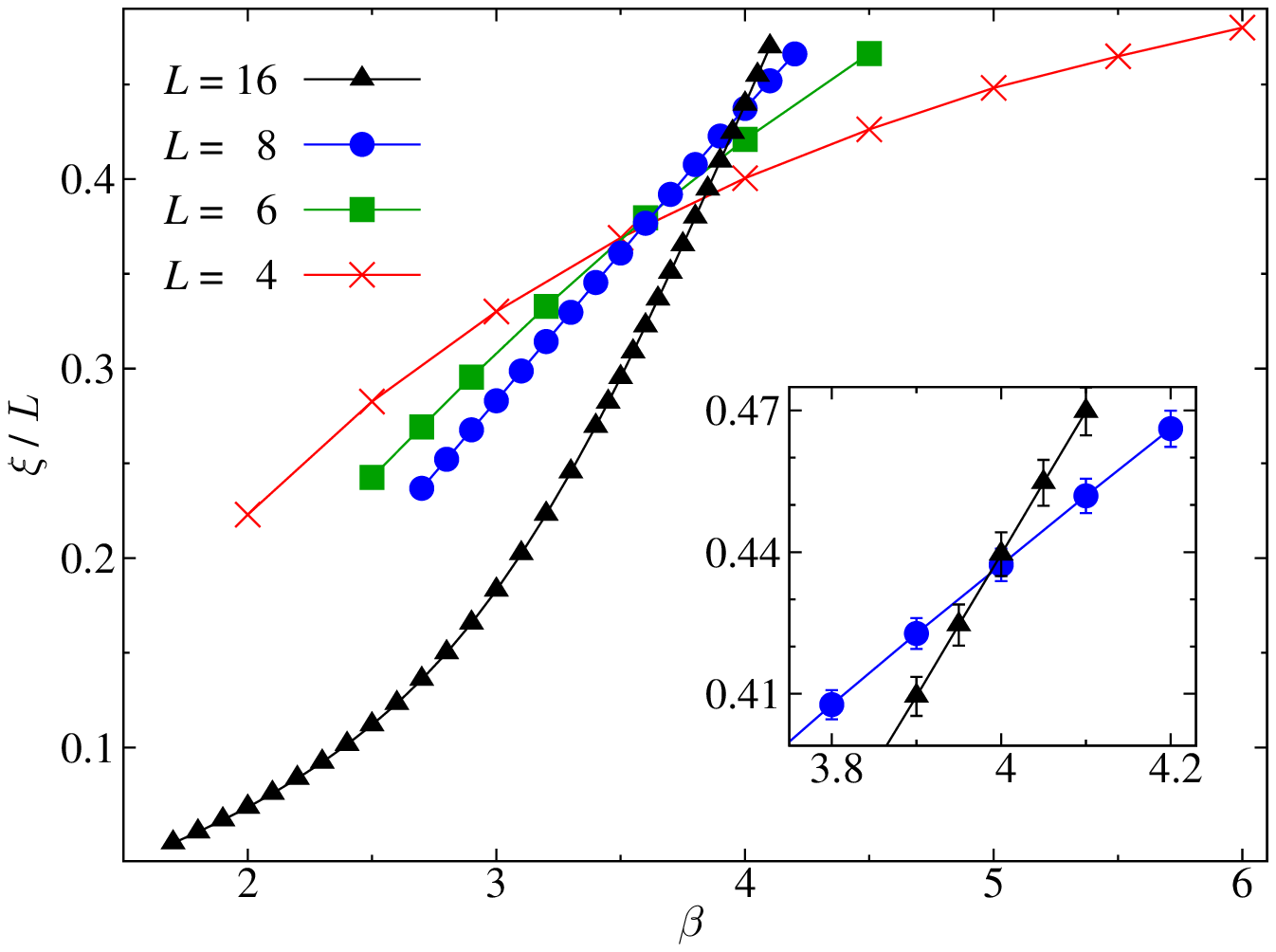}
  \caption{The spin glass correlation length divided by $L$ as a function
    of $\beta$ for $L=4$, $6$, $8$ and $16$.  In the inset we magnify the
    crossing between the $L=8$ and $L=16$ curves.}
  \label{fig:cortes}
\end{figure}

In Fig.~\ref{fig:cortes}, we plot the correlation length (defined in
Eq.~(\ref{eq:xi})) divided by
system size for different lattice
sizes as a function of the temperature. According to
Eq.~(\ref{BETANAIVE}), the data should cross if there is a
transition. Indeed there are clear crossings, indicating a second order
phase transition, though the values of $\beta$ at which the crossings
occur vary with system size.
From Fig.~\ref{fig:cortes} we determined the crossing values
$\beta_{\mathrm{cross}}$ for the pairs of sizes $(4, 8)$ and $(8, 16)$,
see Table~\ref{tab1}.
By computing the 
spin glass susceptibility and the derivative of the correlation length
at these crossing points, we obtain estimates
of the corresponding effective critical exponents, $\eta_q$
and $\nu$, from Eq.~(\ref{QUO}) and also show these results in
Table~\ref{tab1}.
Since we have only data at a discrete set of
temperatures,
we needed an accurate interpolating procedure
to determine the crossing points and the values of other measurables
at these points. 
We chose to fit all
available data with a cubic spline.  To test that our results
are independent of the interpolation procedure we also
implemented a linear interpolation around the crossing point. We 
computed the crossing point and effective exponents with
both procedures, and found agreement within 
the statistical precision of our results.

The two values of $\beta_{\mathrm{cross}}$ shown in Table~\ref{tab1}
are rather different, suggesting large corrections to scaling, i.e.~a
small value for the correction exponent $\omega$, so
we do not have enough information to reliably
compute asymptotic critical exponents.
Nonetheless, from
Table \ref{tab1} we see that the trend of $\eta_q$ with increasing size
is very different from what would be observed in the absence of a
transition for which $\eta_q$ would equal $2$. 
Hence, our numerical
data strongly support the existence of a spin glass phase
transition at finite temperature. 

\begin{table}[h]
  \begin {center}
    \begin{tabular}{|c|c|c|c|c|}  \hline 
      $(L_1, L_2)$ & $\beta_\mathrm{cross}(L_1,L_2)$ &$\nu(L_1,L_2)$ & $\eta_q(L_1,L_2)$ &$ \eta_m(L_1,L_2)$ \\ \hline\hline
      $(4,8)$ & 3.59(4)& 0.83(5) & 0.15(4) & 1.84(3) \\ \hline
      $(8,16)$& 4.00(4) & 0.96(8)& 0.12(6) & 2.06(3)   \\
      \hline 
    \end{tabular}
  \end{center}
  \caption{Results for the critical exponents using the quotient
    method. $(L_1,L_2)$ are the two lattice sizes used and
    $\beta_{\mathrm{cross}}$ is the inverse temperature
    where the two curves of the dimensionless correlation length $\xi/L$ cross
    (see Fig. \ref{fig:cortes}).  The values for $\nu$ and $\eta_q$
    are extracted from measurements involving the overlap $q$,
    whereas $\eta_m$ has been
    computed from the magnetization. These results were obtained with the
    cubic spline interpolating procedure.}
  \label{tab1}
\end{table}

%%%%%%%%%%%%%%%%%%%%%%%%%%%%%%%%%%%%%%%%%%%%%%%%%%%%%%%%%%%%%%%%%%
\subsection{The Magnetization in the Critical
  Region}\label{sub:MAG}

As discussed in the introduction, the standard Potts glass 
studied here 
could undergo a ferromagnetic phase transition at low $T$. This
second transition could bias our analysis
of the spin glass phase by influencing the behavior even 
close to the glass transition temperature (a serious
problem if the two temperature values are very close).  It is therefore
important to investigate
whether there is a region with non-zero spontaneous magnetization
close to the spin glass critical region.

We have therefore computed, using the quotient method,
the growth of the magnetic susceptibility, showing the results in
the last column of
Table \ref{tab1}. The magnetic susceptibility diverges with an exponent
$2-\eta_m$, so $\eta_m\simeq2$ is a clear footprint for the absence of
a magnetic phase transition. For the two largest
lattices we find a value statistically compatible with $2$. Hence we can
safely discard the scenario where a ferromagnetic transition 
appears at $\beta_{\mathrm{cross}}$. We are observing just a glass
transition.

\begin{figure}[h]
 \includegraphics[width=0.7 \columnwidth,angle=270]{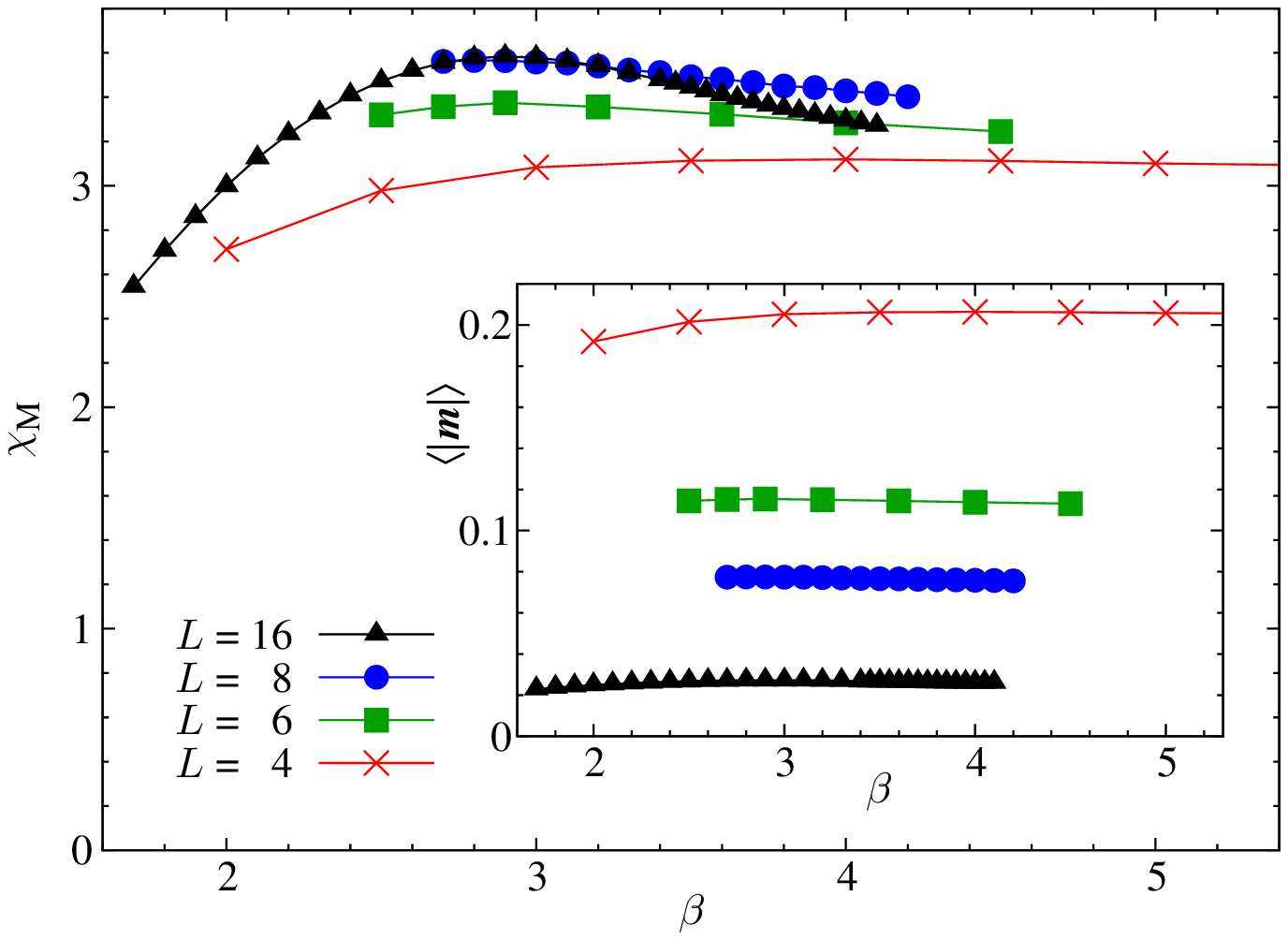}
  \caption{ We show the behavior of the magnetic susceptibility,
    $\chi_{\mathrm{M}}$, versus the inverse temperature. Notice that
    this susceptibility saturates in the critical region. In the inset
    we have plotted the average of the modulus of the magnetization,
    $\overline{<|\mathitbf{m}|>}$, against the temperature: this
    observable behaves as $1/\sqrt{N}$ which clearly signals a
    paramagnetic behavior.}
  \label{fig:mag}
\end{figure}

In order to argue that there is no ferromagnetic transition in the
\textit{whole} temperature range studied, we computed the 
magnetization and susceptibility throughout this range.
In the paramagnetic phase, the magnetization is
random in sign so its modulus
$\overline{<|\mathitbf{m}|>}$ is proportional to $1/\sqrt{N}$, and the
magnetic susceptibility
$ \chi_\mathrm{M} = N \overline{<|\mathitbf{m}|^2>} $
is independent of size.
By contrast, in a ferromagnetic phase,
$\overline{<|\mathitbf{m}|>}$ tends to a positive value at large $N$ 
so $\chi_\mathrm{M}$ diverges proportionally to $N$.

In the main part of
Fig. \ref{fig:mag} we plot $\chi_\mathrm{M}$ versus the inverse of
the temperature. In the glass
pseudo-critical region, $\beta \sim 3.5 - 4$, the two largest lattices
give very similar results, so we recover the result $\eta_m$=2 quoted
in Table \ref{tab1}. Furthermore, at 
\textit{no} temperature does the susceptibility increase
strongly with size. Similarly, the magnetization, shown in the inset of Fig
\ref{fig:mag},  decreases rapidly with size, which also indicates
paramagnetic behavior.

From Fig. \ref{fig:mag} we conclude that there is no ferromagnetic
phase in the region $\beta \in [0,\simeq 4.5]$.

%%%%%%%%%%%%%%%%%%%%%%%%%%%%%%%%%%%%%%%%%%%%%%%%%%%%%%%%%%%%%%%%%%%%%%%%
\section{Conclusions}\label{CONCLUSIONS}

In this study we have numerically explored the
equilibrium behavior of a Potts glass with binary couplings
on large lattices ($L\le 16$ ). A prototype board ($16$ FPGA
processors) of the Janus \cite{IANUS2} optimized computer, using a parallel
tempering algorithm\cite{PT}, has allowed us
to do this for the first time.

By computing the critical exponent associated with
the magnetic susceptibility, and by analyzing the behavior of the
magnetization in the critical region,
we have shown that a paramagnetic-ferromagnetic
phase transition does not occur.
This result is different from 
mean field theory
where, for general $p$, one sees 
both ferromagnetic and spin glass transitions at temperatures which become
equal for $p = 4$. 

We have found and characterized a spin-glass phase transition with
critical exponents $\nu \simeq 1$, $\eta_q \simeq 0.1$ and hence $\beta_q
\simeq 1/2$. In order to extrapolate
these values to the thermodynamic limit,
larger lattice sizes need to be
simulated (which we will try to accomplish in the near future).
The critical exponents computed here are compatible with known values
for other values of $p$.  We note that the exponent $\nu$ decreases with
increasing number of states $p$, since $\nu=2.45(15)$ for $p=2$ (see
Ref.~[\onlinecite{PELI}]) and $\nu= 1.18(5)$ for $p=3$
(see Ref.~[\onlinecite{KLY}]).
The values presented in Table \ref{tab1} for $p=4$
are consistent with this decrease which is expected to end when
$\nu=\tilde{\nu}=2/D=(= 2/3\ \mbox{in}\ D = 3)$, since a finite size
scaling estimate implies that, for a disordered system, the transition
is then first order.\cite{DAFF} Similarly, the exponent $\eta$ grows
with $p$ since $\eta=-0.375(10)$ for $p=2$ and $\eta=0.02(2)$ for $p=3$,
while our estimates in Table \ref{tab1} are larger. 

The hypothesis of a
disordered first order phase transition provides an upper bound
$\eta=1/2$ (since the susceptibility is expected to grow as $L ^{d/2}$).
In the mean field solution of Potts glass, 
second order phase transitions are found for small values ($p \le 4$),
but a first order transition is found\cite{GROSS} 
for $p>4$. An interesting problem for future study is whether
the transition remains second order at large $p$ for short-range spin
glasses in three dimensions or whether,
for a given value of $p>4$, the second order
transition disappears (in a tricritical point) to be replaced by
a first order phase transition at larger $p$. 
\footnote{Notice that the inverse critical temperatures for the $p$-state
  Potts glass model follow well the law: $\beta_c(p) \simeq p$ (e.g.
  $\beta_c(p=2) \simeq 2\times 0.90=1.80$; $\beta_c(p=3) \simeq 2.65$ and in
  this work $\beta_c(p=4) \simeq 4$). If this empirical law is accurate, we
  would expect $\beta_c(p=10) \simeq 10$. This appears to contradict the
  conclusions of Refs.~[\onlinecite{BBK,KLY}], who argued that there is no
  transition for $p=10$. However, we note that virtually all the data in those
  papers is for values of $\beta$ smaller than 10.}

We have studied the standard Potts glass which is expected to be in the same
universality class as the
permutation Potts glass \cite{GLASSY-POTTS}. However, the 
the present state of the art in numerical simulation does not enable us to
confirm this.
In the permutation Potts glass one does not
observe a clear cut phase transition.  Instead of a crossing point 
there is a smooth merging of the curves for different lattice size. This
could indicate transient behavior, i.e.~there is really a phase
transition but it is only visible
on larger lattices, or a Kosterlitz-Thouless like
transition.\cite{POTTS-PERM} For the standard
Potts glass studied here, we find a finite transition as indicated by a
crossing of the correlation length data in Fig.~\ref{fig:cortes}. However, we
note that the crossing point shifts to larger $\beta$, i.e.~smaller $T$, at
larger sizes. It is therefore possible that asymptotic critical behavior could
be quite marginal, as is found in the permutation Potts glass.
Consequently the standard and permutation Potts glass models
\textit{may} be in the same universality class, but larger sizes are
needed to confirm this.

%%%%%%%%%%%%%%%%%%%%%%%%%%%%%%%%%%%%%%%%%%%%%%%%%%%%%%%%%%%%%%%%%
\section*{Acknowledgments}

Janus has been funded by European Union (FEDER) funds, Diputaci\'on General de
Arag\'on (Spain), Microsoft--Italy and Eurotech.  We were partially supported
by MEC (Spain) through contracts TEC2007-64188, FIS2006-08533, FIS2007-60977
and FIS2008-01323, from CAM (Spain) under contract CCG07-UCM/ESP-2532, and
from the Microsoft Prize 2007.  D.~Sciretti acknowledges an FPI fellowship
BFM2003-08532-C03-01 from MEC (Spain). The authors would like to thank the
Ar\'enaire team, especially J\'er\'emie Detrey and Florent de Dinechin for the
VHDL code of the logarithm function.\cite{LOG}

%%%%%%%%%%%%%%%%%%%%%%%%%%%%%%%%%%%%%%%%%%%%%%%%%%%%%%%%%%%%%%%%%%%

\end{document}